\documentstyle[12pt]{article}
\renewcommand{\thesection}{\Roman{section}}
\title{Relativistic effects in the scalar meson dynamics}
\author{R. Kami\'nski, L. Le\'sniak \\Department of Theoretical
Physics,\\
Institute of Nuclear Physics, PL 31-342 Krak\'ow, Poland
\\and\\
J.-P. Maillet\\
Division de Physique Th\'eorique\thanks{Unit\'e de Recherche des
Universit\'es
Paris 11 et Paris 6 associ\'ee au CNRS}, IPN, F-91406 Orsay,\\France}

\begin{document}

\maketitle

\begin{abstract}
   A separable potential formalism is used to describe the
\mbox{$\pi\pi$} and \mbox{$K\overline{K}$} interactions in the
\mbox{I$^{G}$(J$^{PC}$)} = \mbox{0$^{+}$(0$^{++}$)} states in
the energy range from the \mbox{$\pi\pi$} threshold up to 1.4 GeV.
Introduction
of relativistic propagators into a system of Lippmann-Schwinger
equations leads to a very good description of the data ($\chi$$^{2}$
= 0.93 per one degree of freedom). Three poles are found in this
energy region:
\mbox{$f_{0} (500)$}
(\mbox{$M\,=\,506\pm 10$} MeV, \mbox{$\Gamma\,=\,494\pm 5$} MeV),
\mbox{$f_{0} (975)$}
(\mbox{$M\,=\,973\pm$2} MeV, \mbox{$\Gamma\,=\,29\pm$2} MeV) and
\mbox{$f_{0} (1400)$}
(\mbox{$M\,=\,1430\pm$5} MeV, \mbox{$\Gamma\,=\,145\pm 25$} MeV).
The \mbox{$f_{0} (975)$} state can be interpreted as a
\mbox{$K\overline{K}$} bound state. The \mbox{$f_{0} (500)$}
state may be
associated with the often postulated very broad scalar resonance under
the \mbox{$K\overline{K}$} threshold (sometimes called $\sigma$ or
$\epsilon$ meson).
The scattering lengths in the \mbox{$\pi\pi$} and
\mbox{$K\overline{K}$} channels have also been
obtained.
The relativistic approach provides qualitatively new results
(for example the appearance of the \mbox{$f_{0} (500)$} ) in
comparison with
previously used nonrelativistic approach. Interactions in both
channels are attractive and have short range form factors.

\vspace{.5cm}PACS number(s): 14.40.Cs, 13.75.Lb, 12.40.Qq, 11.80.Gw

\end{abstract}

\section{Introduction}

\hspace{.5 cm}
Scalar meson spectroscopy is still far from being well understood
[1--4]. Since it was difficult to explain known properties of the
\mbox{I$^{G}$(J$^{PC}$)} = \mbox{0$^{+}$(0$^{++}$)} states using a
standard
\mbox{$q\bar{q}$} picture, other models treating the four-quark states
\cite{jaffe,achasov84} or meson-meson molecules [7--11] have
been invented. There is also a continuous search \cite{lind,heusch}
for scalar gluonium states which can be mixed with the quark states of
the same quantum numbers. The nature of the \mbox{$f_{0} (975)$} and
\mbox{a$_{0}(980)$} mesons is very controversial [14--17].
Closeness of
their almost degenerated masses to the \mbox{$K\overline{K}$}
threshold energy constitutes an
argument towards their interpretation as virtual bound
\mbox{$K\overline{K}$} states
which are unstable due to the open \mbox{$\pi\pi$} channel
[10, 18--21].

   In this article we extend the coupled channel formalism of
Refs~\cite{cann89,cann92}
including some relativistic effects in both the \mbox{$\pi\pi$} and
\mbox{$K\overline{K}$} decay channels of the scalar mesons.
We use the relativistic propagators
and the separable potentials in the Lippmann--Schwinger formalism.
Such an approach
has already been applied in the analysis of the pion-nucleon
amplitudes
\cite{lond74} or in the studies of the \mbox{$\Lambda(1405)$}
resonance structure \cite{siegel88,fink90}.
Our aim is to describe quantitatively the isoscalar $s$-wave
\mbox{$\pi\pi$} and the \mbox{$K\overline{K}$} scattering data in a
wide energy range starting from the \mbox{$\pi\pi$} threshold up to
1.4 GeV.
The parameters of the meson-meson interactions are fitted to the data
and the
$S$-matrix structure is analysed in order to extract information about
the
scalar resonances in that energy region. This is a new step in
comparison with
\cite{cann89} where some pole structure of the $S$-matrix has been
postulated
and the $\chi^2$ tests have not been done. We observe that the
relativistic
effects are important not only in the \mbox{$\pi\pi$} channel but
also in the \mbox{$K\overline{K}$} channel even near the threshold.
This fact may have important implications, for example
the \mbox{$f_{0} (975)$} meson, interpreted as a
\mbox{$K\overline{K}$} molecule, can still have a relatively
small radius (less than 1 fm). Consequently, the predictions of the
radiative
$\phi$ decay into the \mbox{\mbox{$K\overline{K}$} $\gamma$} system
via \mbox{$\phi\rightarrow \mbox{$f_{0} (975)$} \gamma$} have to be
substantially influenced \cite{close92}.
The importance of the relativistic effects on the two photon decays
of the \mbox{J$^{PC}$} = \mbox{0$^{++}$} and \mbox{2$^{++}$} states
has also been recently stressed \cite{li91}.

   This paper is organized as follows. In Sect. II we define our
formalism.
Sect. III is devoted to an analysis of the single \mbox{$\pi\pi$}
channel essentially up to the
energy of about 700 MeV. The properties of the
\mbox{$K\overline{K}$} channel interaction in the
decoupled case are discussed in Sect. IV. Both the nonrelativistic
and relativistic
cases are treated. In Sect. V we analyse the interactions in the
coupled \mbox{$\pi\pi$}
and \mbox{$K\overline{K}$} channels. A comparison with the data is
performed and the $S$-matrix structure is studied.
In Sect. VI we summarize the main results and outline some
perspectives.

\section{The Formalism}

\hspace{.5 cm}
We describe the \mbox{$\pi\pi$} and \mbox{$K\overline{K}$} isoscalar
$s$-wave interactions in the framework
of the coupled channel Lippmann-Schwinger equations \cite{lond74}.
The scattering
amplitude $T$ satisfies the following equation in the momentum space:
\begin{eqnarray}
< {\bf p} \mid T \mid {\bf q} > & =& < {\bf p} \mid V \mid {\bf q} >
+ \nonumber\\
& & \int \frac{d^{3}s}{(2\pi)^{3}} < {\bf p} \mid V \mid {\bf s} >
< {\bf s} \mid G \mid {\bf s} >< {\bf s} \mid T \mid {\bf q} >,
\label{l_s}
\end{eqnarray} \\
where $V, G, T $ are 2$\times$2 matrices (label 1 denotes the
\mbox{$K\overline{K}$}
channel and label 2 -- the \mbox{$\pi\pi$} channel), $V$ is the
interaction
matrix and $G$ is the diagonal matrix of propagators written in the
center of mass system:
\begin{equation}
 < {\bf s} \mid G_{ij} \mid {\bf s} > = G_{i}(s)\,\delta_{ij}
\;\;\;\;\;(i,j=1,2),
\label{G_def}
\end{equation}
and $G_{i}^{-1}(s) = E-2E_{i}(s)+i\epsilon,\,\epsilon \longrightarrow
0(+)$.
In Eq.~(\ref{G_def}) ${\bf s}$ is the relative momentum, $E$ is the
total energy and $E_{i}(s)=\sqrt{s^{2}+m_{i}^{2}}$
are the relativistic energies; $m_{1}$=495.69 MeV, $m_{2}$=137.27 MeV
are the average kaon and pion masses.

   A very convenient parameterization of the interaction matrix is a
separable
potential form. In the \mbox{$K\overline{K}$} channel we use the
simplest rank--one potential in the momentum space
\begin{equation}
< {\bf p} \mid V_{11} \mid {\bf q} >=
\mbox{$\lambda_{11}$}\,g_{1}(p)g_{1}(q)
\label{k_pot}
\end{equation}
and in the \mbox{$\pi\pi$} channel a rank--two potential
\begin{equation}
< {\bf p} \mid V_{22} \mid {\bf q} > =
\mbox{$\lambda_{22}$}\,g_{2}(p)g_{2}(q)+
\mbox{$\lambda_{33}$}\,g_{3}(p)g_{3}(q).
\label{p_pot}
\end{equation}
The transition potential matrix elements read
\begin{equation}
< {\bf p} \mid V_{12} \mid {\bf q} >=
< {\bf q} \mid V_{21} \mid {\bf p} >=
\mbox{$\lambda_{12}$}\,g_{1}(p)g_{2}(q)+
\mbox{$\lambda_{13}$}\,g_{1}(p)g_{3}(q).
\label{v12_pot}
\end{equation}
In Eqs.~(\ref{k_pot})--(\ref{v12_pot}) \mbox{$\lambda_{ik}$}
$(i,k=1,2,3)$ are the coupling
constants and $g_{i}$ are the form factors which we have chosen in
the Yamaguchi form \cite{yam54}:
\begin{equation}
g_{i}(p)=\sqrt{\frac{4\pi}{m_{i}}}\,\frac{1}{p^{2}+\beta_{i}^{2}},
\label{g_def}
\end{equation}
where
\begin{equation}
m_{i}=\left\{ \begin{array}{ll}
              m_{K} & \mbox{if \hspace{.1cm} i=1} \\
              m_{\pi} & \mbox{if \hspace{.1cm} i=2,3}
              \end{array}
      \right.
\label{mi_def}
\end{equation}
and $\beta_{i}$ are the form factor range parameters. The potentials
(\ref{k_pot}--\ref{v12_pot}) are nonlocal.
After this choice the potential matrix has eight
parameters (five coupling constants and three range parameters)
which should
be fixed by a comparison of the theoretical amplitudes with
experimental data.

   The separable form of the interactions enables us to write the
following scattering matrix elements:
\begin{eqnarray}
<{\bf p}\mid T_{11} \mid {\bf q}>&=& g_{1}(p)\,t_{11}\,g_{1}(q),
\label{T11}\\
<{\bf p}\mid T_{22} \mid {\bf q}>&=& g_{2}(p)\,t_{22}\,g_{2}(q)+
g_{2}(p)\,t_{23}\,g_{3}(q) \nonumber\\
&+& g_{3}(p)\,t_{32}\,g_{2}(q)+g_{3}(p)\,t_{33}\,g_{3}(q),
\label{T22}\\
\vspace{.1cm}
<{\bf p}\mid T_{12} \mid {\bf q}>&=&<{\bf q}\mid T_{21} \mid {\bf p}>=
g_{1}(p)\,t_{12}\,g_{2}(q)+g_{1}(p)\,t_{13}\,g_{3}(q), \label{T12}
\end{eqnarray}
where $t_{kl}$ are energy dependent reduced amplitudes ($k,l=1,2,3$).
In Eq.~(\ref{T11}) $T_{11}$ denotes the \mbox{$K\overline{K}$} elastic
 scattering
amplitude, $T_{22}$ in Eq.~(\ref{T22})
is the \mbox{$\pi\pi$} elastic scattering amplitude while
$T_{12}$ and $T_{21}$ are the transition
\mbox{$\mbox{$K\overline{K}$}\rightarrow\mbox{$\pi\pi$}$}
and \mbox{$\mbox{$\pi\pi$}\rightarrow\mbox{$K\overline{K}$}$}
amplitudes. The system (\ref{l_s}) of
the coupled integral equations satisfied by the $T_{kl}$ elements
leads to a set of algebraic equations for the reduced amplitudes
$t_{kl}$ written in the $3\times3$ matrix form:
\begin{equation}
t=\lambda+\lambda\,I\,t.  \label{t}
\end{equation}
In Eq.~(\ref{t}) $\lambda$ is the symmetric $3\times3$ matrix of the
coupling constants
\begin{equation}
\label{lda_def}
 \lambda = \left( \begin{array}{ccc}
		\mbox{$\lambda_{11}$} & \mbox{$\lambda_{12}$} &
\mbox{$\lambda_{13}$}\\
	\mbox{$\lambda_{12}$} & \mbox{$\lambda_{22}$} & 0\\
	\mbox{$\lambda_{13}$} & 0	  & \mbox{$\lambda_{33}$}
	      \end{array}
		\right)
\end{equation}
and $I$ is the following symmetric matrix
\begin{equation}
\label{I_def}
 I = \left( \begin{array}{ccc}
		\mbox{$I_{11}$} &  0    &   0\\
		0     & \mbox{$I_{22}$} & \mbox{$I_{23}$}\\
		0     & \mbox{$I_{23}$} & \mbox{$I_{33}$}
	      \end{array}
		\right)
\end{equation}
consisting of the integrals
\begin{equation}
\mbox{$I_{11}$}=
\int \frac{d^{3}s}{(2\pi)^{3}}\,g_{1}(s)G_{1}(s)g_{1}(s)
\label{I_ii}
\end{equation}
and
\begin{equation}
I_{kl}=\int \frac{d^{3}s}{(2\pi)^{3}}\,g_{k}(s)G_{2}(s)g_{l}(s),
\;\;\;(k,l=2,3).
 \label{I_kl}
\end{equation}
In Eqs.~(\ref{g_def}), (\ref{lda_def}) and (\ref{I_def}) label~1
refers to the
\mbox{$K\overline{K}$} channel and labels 2 and 3 to the
\mbox{$\pi\pi$} channel in which the interaction
potential contains two terms. The matrices $t$ and $I$ are functions
of the
energy $E$ (see Eq.~(\ref{G_def})).

   A solution of Eq.~(\ref{t}) is straightforward:
\begin{equation}
t=(1-\lambda I)^{-1}\lambda \label{t_sol}
\end{equation}
and the resulting $t$-matrix is symmetric ($t_{kl}=t_{lk}$) since the
matrices $\lambda$ and $I$ are symmetric.
Explicit expressions for the matrix elements $t_{kl}$ are given in the
Appendix. All the functions $t_{kl}$ are inversely
proportional to the Jost function
\begin{equation}
D(E)=det(1-\lambda I). \label{det}
\end{equation}

   Let us notice that the coupling constants \mbox{$\lambda_{ij}$}
defined by
(\ref{k_pot})--(\ref{v12_pot}) have the dimension (MeV/c)$^{3}$ and
the integrals (\ref{I_ii}--\ref{I_kl}) -- the dimension (MeV)$^{-3}$.
For simplicity in further calculations we will use the dimensionless
coupling constants defined as
\begin{equation}
\mbox{$\Lambda_{ij}$}=
\frac{\mbox{$\lambda_{ij}$}}{2(\beta_{i}\beta_{j})^{3/2}}
\label{dim_less_lda}
\end{equation}
and redefined integrals
\begin{equation}
J_{ij}=2(\beta_{i}\beta_{j})^{3/2}\,I_{ij}. \label{J_def}
\end{equation}

   If the energy $E$ is higher than the \mbox{$K\overline{K}$}
threshold mass we can
relate \mbox{(compare \cite{lond74})} the on-shell scattering matrix
elements $T_{kl}(k_{1},k_{2})$
to the \mbox{$S$-matrix} elements
\begin{eqnarray}
S_{11}&=& 1-\frac{ik_{1}E_{1}(k_{1})}{2\pi}\,T_{11}(k_{1},k_{2}),
\label{S11}\\
S_{22}&=& 1-\frac{ik_{2}E_{2}(k_{2})}{2\pi}\,T_{22}(k_{1},k_{2}),
\label{S22}\\
S_{12}&=& S_{21}=-\frac{i}{2\pi}\sqrt{k_{1}E_{1}(k_{1})k_{2}E_{2}
(k_{2})}\,T_{12}(k_{1},k_{2}), \label{S12}
\end{eqnarray}
where the \mbox{$K\overline{K}$} channel and \mbox{$\pi\pi$} channel
momenta $k_{1}$ and $k_{2}$ are defined by
the energy conservation condition:
\begin{equation}
E=2\sqrt{k_{1}^{2}+m_{K}^{2}}=2\sqrt{k_{2}^{2}+m_{\pi}^{2}}.
\label{en}
\end{equation}
This relation allows one to write the functions $I_{11}(E)$ and
$I_{kl}(E)$
appearing in Eqs.~(\ref{I_ii}) and (\ref{I_kl}) as functions
$I_{11}(k_{1})$ and
$I_{kl}(k_{2})$. Similarly the Jost function $D(E)$ given by
(\ref{det}) can be
expressed as a function of two related variables
\begin{equation}
D(k_{1},k_{2})=D_{K}(k_{1})D_{\mbox{$\pi$}}(k_{2})-C(k_{1},k_{2}),
\label{DKPC}
\end{equation}
where
\begin{eqnarray}
D_{K}(k{_1})&=&1-\Lambda_{11}\,J_{11}(k_{1}), \label{K}\\
D_{\mbox{$\pi$}}(k_{2})&=&1-\Lambda_{22}\,J_{22}(k_{2})-
\Lambda_{33}\,J_{33}(k_{2})+\Lambda_{22}\Lambda_{33}\,d(k_{2}),
\label{P}\\
C(k_{1},k_{2})&=
&\Lambda_{12}\,J_{11}(k_{1})[\Lambda_{12}\,J_{22}(k_{2})+
\Lambda_{13}\,J_{23}(k_{2})-
\Lambda_{12}\Lambda_{33}\,d(k_{2})] \nonumber \\
&+& \Lambda_{13}\,J_{11}(k_{1})[\Lambda_{12}\,J_{23}(k_{2})+
\Lambda_{13}\,J_{33}(k_{2})-\Lambda_{13}\Lambda_{22}\,d(k_{2})]
\label{C}
\end{eqnarray}
and
\begin{equation}
d(k_{2})=J_{22}(k_{2})\,J_{33}(k_{2})-J_{23}^{2}(k_{2}). \label{d}
\end{equation}
In Eq.~(\ref{K}) $\!D_{K}(k_{1})$ is a part of the Jost function
corresponding to the
interaction in the single \mbox{$K\overline{K}$} channel,
$D_{\mbox{$\pi$}}(k_{2})$ is a similar part in
the \mbox{$\pi\pi$} channel and $C(k_{1},k_{2})$ comes from the
interchannel coupling.

   Using the Jost function $D$ we can also express the diagonal
$S$-matrix elements as the ratios:
\begin{eqnarray}
S_{11}&=&\frac{D(-k_{1},k_{2})}{D(k_{1},k_{2})}, \label{SD11}\\
S_{22}&=&\frac{D(k_{1},-k_{2})}{D(k_{1},k_{2})}, \label{SD22}
\end{eqnarray}
and the nondiagonal matrix element satisfies the equation
\begin{equation}
S_{11}S_{22}-S_{12}^{2}=\frac{D(-k_{1},-k_{2})}{D(k_{1},k_{2})}.
\label{SD12}
\end{equation}
Above the \mbox{$K\overline{K}$} threshold the $S$-matrix can be
parameterized in terms of the
inelasticity parameter $\eta$ and the phase shifts $\delta_{\pi\pi}$
and $\delta_{K\bar{K}}$:
\begin{equation}
S = \left( \begin{array}{cc}
\eta e^{2i\delta_{K\overline{K}}}  &
i\sqrt{1-\eta^{2}}e^{i(\delta_{\pi\pi}+\delta_{K\overline{K}})}\\
i\sqrt{1-\eta^{2}}e^{i(\delta_{\pi\pi}+\delta_{K\overline{K}})} &
\eta e^{2i\delta_{\pi\pi}}
	      \end{array}
		\right).
\end{equation}
If the energy $E$ is lower than $2m_{K}$ then we can still use
Eqs.~(\ref{S22})
and (\ref{SD22}), provided $k_{1}$ is purely imaginary
$k_{1}=i\sqrt{m_{K}^{2}-E^{2}/4}$.
In this case the inelasticity parameter $\eta\equiv 1$
since the \mbox{$K\overline{K}$} channel is closed.

   Energy dependence of the experimentally measured quantities
$\delta_{\pi\pi}$,
$\delta_{K\bar{K}}$ and $\eta$ is closely related to the analytical
structure
of the Jost function in the complex planes of the $k_{1}$ and $k_{2}$
momenta.

\section{$\pi\pi$ channel interactions}

\hspace{.5cm}
In this section we consider interactions in the single \mbox{$\pi\pi$}
channel
i.e. without couplings to the \mbox{$K\overline{K}$} channel
($\mbox{$\lambda_{12}$}=\mbox{$\lambda_{13}$}=0$).
We study the
energy range from the \mbox{$\pi\pi$} threshold up to about 700 MeV,
where in the first
step the influence of the coupling to the \mbox{$K\overline{K}$}
channel can be neglected.
The aim of such an analysis is an estimation of the \mbox{$\pi\pi$}
potential parameters
and investigation of the pole structure of the \mbox{$\pi\pi$}
scattering function $S_{22}$.
Two experimental facts can be quantitatively described in such an
approach:
the first is a positive value of the \mbox{$\pi\pi$} $s$--wave
scattering length in
the $I=0$ \mbox{$s$--state}, and the second is a systematic increase of
the \mbox{$\pi\pi$} phase shifts from 0
degrees at the \mbox{$\pi\pi$} threshold up to about 70 degrees at
the energy of 700 MeV.
Experimental data indicate the existence of the \mbox{$f_{0} (1400)$}
meson.
Although the \mbox{$f_{0} (1400)$} mass is higher than the energy
limit under consideration, this resonance should be
included in our analysis because of its large coupling to the
\mbox{$\pi\pi$} channel and a large width.

   At the beginning let us discuss a case when the propagator $G_{2}$
(see Eq.~(\ref{G_def})) has the nonrelativistic form
$G_{2}^{-1}(s)=(k_{2}^{2}-s^{2})/m_{2}+i\epsilon$.
We start by assuming that in the potential $V_{22}$ in
Eq.~(\ref{p_pot}) only the first term, i.e. the coupling constant
$\mbox{$\lambda_{33}$}=0$, is present.
Then the Jost function $D_{\mbox{$\pi$}}(k_{2})$ can be expressed as
(see Eqs.~(\ref{P}) and (\ref{aI_ii_n}) in Appendix)
\begin{equation}
D_{\mbox{$\pi$}}(k_{2})=1+\frac{\Lambda_{22}}{(1-ia_{2})^{2}},
\label{d_non}
\end{equation}
where $a_{2}=k_{2}/\beta_{2}$. Its phase is directly related to the
phase shift $\delta_{\pi\pi}$ by
\begin{equation}
D_{\mbox{$\pi$}}(k_{2})=\mid D_{\mbox{$\pi$}}(k_{2})
\mid e^{-i\delta_{\pi\pi}}.
\label{s_pp}
\end{equation}
Knowing that the scattering length
\begin{equation}
a_{\pi\pi}= \lim_{k_{2} \longrightarrow 0} \delta_{\pi\pi}/k_{2},
\label{low_en_apr}
\end{equation}
we can evaluate a low energy limit of the Jost function (\ref{d_non}).
As a result we obtain
\begin{equation}
a_{\pi\pi}=-\frac{2\Lambda_{22}}{\beta_{2}(1+\Lambda_{22})}.
\label{scatt_n}
\end{equation}
The positivity of the experimental $a_{\pi\pi}$ value imposes the
following conditions on the coupling constant \mbox{$\Lambda_{22}$}:
\begin{equation}
-1<\Lambda_{22}<0.  \label{l2cond_n}
\end{equation}
The analytical structure of the Jost function in Eq.~(\ref{d_non})
is very simple:
it has a double pole at $k_{2}=-i\beta_{2}$ and two single roots at
\mbox{$k_{2}=i\beta_{2}(\pm\sqrt{-\Lambda_{2}}-1)$}.
For the \mbox{$\Lambda_{22}$} satisfying the inequalities
(\ref{l2cond_n}) both zeros lie on the negative part of the imaginary
$k_{2}$--axis. Since the matrix element $S_{22}$ is given by a ratio
$D_{\mbox{$\pi$}}(-k_{2})/D_{\mbox{$\pi$}}(k_{2})$, its poles coincide
with the Jost function zeros
of the denominator except of the numerator double pole at
$k_{2}=i\beta_{2}$.
The existence of the \mbox{$f_{0} (1400)$} meson must manifest itself
by the presence
of two complex conjugated $S$--matrix poles lying off the imaginary
$k_{2}$--axis.
It has been shown in \cite{cann89} that the introduction of the second
term in
the potential $V_{\pi\pi}$ permits one to obtain such a pair of poles
and, in
addition, a positive value of the scattering length. The coupling
constants
\mbox{$\Lambda_{22}$} and \mbox{$\Lambda_{33}$} in \cite{cann89} had
opposite signs.
Although it was not shown there, we can prove that in the case when
the \mbox{$\pi\pi$}
potential contains in Eq.~(\ref{p_pot}) two terms, when there is no
coupling to
the \mbox{$K\overline{K}$} channel, and the scattering length is
positive, the
two other poles of the $S_{22}$ can lie only on the imaginary axis.

   Next we discuss the relativistic form of the propagator $G_{2}$ and
investigate
the analytical structure of the matrix element $S_{\mbox{$\pi\pi$}}$
calculated for a single term in $V_{\pi\pi}$
($\mbox{$\lambda_{33}$}=0$).
The situation now is quite different:
the $S_{22}$ function has a pair of poles off the imaginary axis and
simultaneously
the condition $a_{\pi\pi}>0$ can be fulfilled. The expression for the
scattering length in the relativistic case is
\begin{equation}
a_{\pi\pi}=-\frac{2\Lambda_{22}}{\beta_{2} \left\{ 1+\Lambda_{22}
\left[ \mbox{$\frac{1}{2}$}
+\frac{1}{\mbox{$\pi$}}\left(\frac{\mbox{$\beta_{2}$}}
{\mbox{$m_{2}$}}+
\frac{\mbox{$m_{2}$}}
{\mbox{$\beta_{2}$}}\,d_{2}\right)\right]\right\}},  \label{scatt_r}
\end{equation}
where $d_{2}$ is a constant defined in Appendix by (\ref{aabdi}) and
(\ref{aF_res}).
In order to get $a_{\pi\pi}>0$ the coupling
constant \mbox{$\Lambda_{22}$} must satisfy the following conditions:
\begin{equation}
-\left[\frac{1}{2}+\frac{1}{\pi}\left(\frac{\beta_{2}}{m_{2}}+
\frac{m_{2}}{\beta_{2}}\,d_{2}\right)\right]^{-1}<\Lambda_{22}<0.
\label{l2cond_r}
\end{equation}
Negative sign of the coupling constant \mbox{$\Lambda_{22}$} means
that the interaction is attractive.

   Using (\ref{s_pp}) we have compared our theoretical predictions
for the \mbox{$\pi\pi$}
phase shifts with the experimental data in the energy range from the
\mbox{$\pi\pi$}
threshold up to 700 MeV. We have used the data from the $K_{e4}$
decay
\cite{rosselet} (5 points) and from the reaction
$\pi^{-}p\rightarrow\pi^{-}\pi^{+}n$ [29--31]
(16 points up to 700 MeV).
We have performed the $\chi^{2}$ test using the CERN program MINUIT
and as the output parameters we obtained $\Lambda_{22}=-0.213$ and
$\beta_{2}=1239$ MeV.
For these parameters $\chi^{2}=1.13$ per one degree of freedom.
The calculated
scattering length $a_{\pi\pi}=0.18\, m_{\pi}^{-1}$ can be compared
with experimental values
$(0.28\pm 0.05)\, m_{\pi}^{-1}$ \cite{rosselet},
$(0.24\pm 0.09)\, m_{\pi}^{-1}$ \cite{belkov} and $(0.207\pm 0.028)\,
m_{\pi}^{-1}$ \cite{lowe}.
In a combined analysis \cite{burkhardt} of the $\pi N \rightarrow
\pi\pi N$
data the value $a_{\pi\pi}=(0.197\pm 0.01)\, m_{\pi}^{-1}$ is quoted.
The scattering $S$--matrix function $S_{22}$ has only one pair of
the complex conjugated poles at the energy $E=491\pm i246$ MeV.
It is clear that it cannot be related to the \mbox{$f_{0} (1400)$}
meson which
according to the Particle Data Group \cite{pdg92} has the energy
$1400-i(75\div 200)$ MeV.
In order to include the \mbox{$f_{0} (1400)$} state in a further
analysis we have added a
second term to the \mbox{$\pi\pi$} potential and the Jost function
$D_{\mbox{$\pi$}}(k_{2})$
obtained the form (\ref{P}). Then in the complex momentum plane we
have found a
second pair of the complex conjugated poles of the $S_{22}$ function
which can
be related to the \mbox{$f_{0} (1400)$} meson. In the numerical
calculations of the pole
positions we must take into account the energy range above 1 GeV and
include the coupling to the \mbox{$K\overline{K}$} channel.
Therefore a  full discussion
concerning the \mbox{$f_{0} (1400)$}
meson is presented in Section V. It is worth--while to note, however,
that
contrary to the nonrelativistic case, both coupling constants
\mbox{$\Lambda_{22}$} and \mbox{$\Lambda_{33}$} have received negative
signs.

   The appearance of a new $S_{22}$ pole, in comparison with the
nonrelativistic case, at an energy about 500 MeV can be related to
the existence of the so--called
$\sigma$ or $\epsilon$ meson which has been often postulated both
experimentally and theoretically (see references in \cite{efimov}).
The mass of this meson varies between about 500 MeV and 1000 MeV,
while the width even more~--~between 300 MeV and 1000 MeV. Various
models of the $\sigma$
meson structure have been formulated describing it as normal
$q\overline{q}$,
$qq\overline{q}\overline{q}$ or gluon--gluon states.
Such a low mass scalar meson also appeared in the so--called
$\sigma$--model
and in other field theoretical models (see \cite{martin} and
references cited
therein) in which its mass had a value of about
500 MeV and its large width varied from 300 MeV to more than 500 MeV.
The exchange of the $\sigma$ meson has been used in the Bonn model of
the nucleon--nucleon interactions \cite{holinde87}. It has been found
that the $\sigma$ meson exchange can effectively describe the
contribution of two
correlated $s$--wave pions in the isospin 0 state.
The $\sigma$ mass and width have been also recently discussed
in the Nambu-Jona-Lasinio model \cite{nakayama}.
An experimental measurement of the $\sigma$ parameters is very
difficult because of its large width [35--39].
This meson does not appear as a typical Breit--Wigner peak and in the
partial wave
analyses it can easily be interpreted as a background. The last time
when the $\sigma$ meson was included in the Particle Data Group
Tables was 1974 with $M_{\sigma}\leq 700$ MeV and
$\Gamma_{\sigma}\geq 600$ MeV (and it was called $\epsilon$)
\cite{pdg74}.

   In the isospin 2, spin 0 \mbox{$\pi\pi$} channel one does
not expect a resonant behaviour of the scattering amplitude.
Using a simple rank-one potential of
the Yamaguchi form we have obtained a very good description of the
corresponding \mbox{$\pi\pi$} phase shifts up to about 1 GeV.
No poles of
the I=2 scattering amplitude have been found in the momentum range
of $Re k_2$ between 0 and 4 GeV and $Im k_2$ between 0 and -4 GeV.
This fact, which confirms the above expectation, is in
contrast to the situation in the isospin 0 channel in which we do
find resonances.

\section{\mbox{$K\overline{K}$} interactions}
\hspace{.5cm}
Let us now discuss the \mbox{$K\overline{K}$} interactions without
coupling to the \mbox{$\pi\pi$} channel.
We shall be especially interested in the energy region close to the
\mbox{$K\overline{K}$} threshold near 1 GeV. In this paper we do not
distinguish the $K^{0}\overline{K^{0}}$ and
$K^{+}K^{-}$ thresholds so we use their average energy value
$E_{th}=2m_{1}=991.38$ MeV. The \mbox{$f_{0} (975)$} meson energy is
very close to this value, so it is
quite natural to interpret this state as a bound
\mbox{$K\overline{K}$} state with a binding
energy of about 16 MeV. We do not, however, postulate the existence
of the bound
state from the beginning. As we shall see in the next chapter, the
parameters of
the \mbox{$K\overline{K}$} interaction will be fixed by a global fit
to the \mbox{$\pi\pi$} and \mbox{$K\overline{K}$} data.
Here we discuss different conditions for the existence of the
particular \mbox{$K\overline{K}$} structures:
bound states, antibound states (sometimes called virtual states
\cite{newton}), or resonances. For each case we shall examine effects
of the relativistic
propagation of kaons and their influence on some observables, like
scattering length or phase shifts. A very good method to distinguish
the three types of the
\mbox{$K\overline{K}$} states is to look at the positions of the
$S$--matrix poles corresponding
to the Jost function zeros in the complex $k_{1}$--plane
(Eq.~\ref{SD11}).
If the zero lies on the positive imaginary axis ($k_{1}$=$i\alpha_{1},
\alpha_{1}>0$),
then a bound state exists. The antibound state corresponds to a zero
lying on the negative part of imaginary axis ($\alpha_{1}<0$).
If there are two
complex conjugated zeros in the lower half plane then there is a
resonance in the \mbox{$K\overline{K}$} channel.

   At first we analyse the most important features of the bound state.
Its binding energy can be defined as
\begin{equation}
E_{B}=2m_{K}-2\sqrt{m_{K}^{2}-\alpha_{1}^{2}} \hspace{.8cm} \mbox{for}
\hspace{.5cm} \alpha_{1}<m_{K}.
\label{EB}
\end{equation}
The Jost function $D_{K}(k_{1})$ in Eq.~(\ref{K}) vanishes if
\begin{equation}
\Lambda_{11}=1/Re[J_{11}(i\alpha_{1})] \label{1/re(ialfa)}
\end{equation}
since $Im[J_{11}(i\alpha_{1})]\equiv 0$. The coupling constant depends
on the range
parameter \mbox{$\beta_{1}$} which can be chosen regardless of the
binding energy $E_{B}$.
In the nonrelativistic case we approximate the
propagator $G_{1}$ in Eq.~(\ref{G_def}) by
\begin{equation}
G_{1}(s)=\frac{m_{K}}{k_{1}^{2}-s^{2}}. \label{G_non}
\end{equation}
Then according to Eq.~(\ref{aI_ii_n}) the dimensionless
\mbox{$K\overline{K}$} coupling constant is particularly simple
\begin{equation}
\Lambda_{11}=-\left( 1+\frac{\alpha_{1}}{\beta_{1}} \right)^{2}.
\label{lda_bound}
\end{equation}
The \mbox{$K\overline{K}$} force is attractive and the coupling
constant is negative and smaller
than $-1$. For the relativistic propagator we have numerically
verified (see also
(Eq.~\ref{aI_ii}))
that the \mbox{$\beta_{1}$}--dependence of \mbox{$\Lambda_{11}$}
is very similar to the one in the
nonrelativistic case for the small values of \mbox{$\beta_{1}$}
up to about 200 MeV.
For larger \mbox{$\beta_{1}$} the absolute value of the
"relativistic" coupling constant
is smaller than the coupling constant in the nonrelativistic limit.
If $\beta_{1}\longrightarrow \infty$, then nonrelativistically
$\Lambda_{11} \longrightarrow -1$,
while in the relativistic case $\Lambda_{11}\longrightarrow 0$ as
$-\pi m_{K}/\beta_{1}$
(see Eqs.~(\ref{aI_ii}) and (\ref{1/re(ialfa)})). This is the first
indication that in the limit of large \mbox{$\beta_{1}$} the
relativistic effects might be important.

   In the absence of the coupling to the \mbox{$\pi\pi$} channel
we can use similar
relation between the \mbox{$K\overline{K}$} scattering phase shift
$\delta_{K\overline{K}}$ and the phase
of the Jost function $D_{K}(k_{1})$ as in Eq.~(\ref{s_pp})  for
the \mbox{$\pi\pi$} channel. In the limit \mbox{$k_{1}
\longrightarrow 0$} the \mbox{$K\overline{K}$} scattering length is
given by \mbox{$a_{K\overline{K}}\approx
tg(\delta_{K\overline{K}})/k_{1}$}. The energy dependence of the
phase shifts can be directly calculated using Eqs.~(\ref{en}),
(\ref{K}),
(\ref{aI_ii}) or (\ref{aI_ii_n}) for a specific choice of the bound
state
energy at a given value of the range parameter \mbox{$\beta_{1}$}.
This is illustrated in
Fig.~1a for two values of \mbox{$\beta_{1}$}=150 MeV and 2000 MeV.
At the \mbox{$K\overline{K}$} threshold the value of the phase
shift is $\pi$
(conventionally assumed for a bound state) and monotonically
decreases to zero as the energy increases. Again we
see a very important difference between the relativistic and
nonrelativistic
expressions for large \mbox{$\beta_{1}$} values ($\beta_{1}>m_{K}$).
At $\beta_{1}=2000$ MeV and
$E=1400$ MeV the difference is as large as $30^{\circ}$. Let us also
notice that
at higher \mbox{$\beta_{1}$} the phase shift decrease with energy is
much less steep than at lower
$\beta_{1}=150$ MeV. This decrease at the \mbox{$K\overline{K}$}
threshold is governed by the
negative value of the \mbox{$K\overline{K}$} scattering length.
The general expression for the
\mbox{$K\overline{K}$} scattering length in the relativistic case is
given by Eq.~(\ref{scatt_r})
provided we substitute in it the \mbox{$K\overline{K}$} channel
parameters $m_{1}$, \mbox{$\beta_{1}$} and \mbox{$\Lambda_{11}$}.

   The values of the scattering lengths and the coupling constants
are given in
Table~\ref{tab_kk_sc_coupl}. As indicated in the last column, the
relativistic
corrections at the \mbox{$K\overline{K}$} threshold amount to 13\%
if the value \mbox{$\beta_{1}$}
is as large as 2000 MeV, while at low \mbox{$\beta_{1}$}
$\leq$ 150 MeV they are smaller than 1\%.

   Next we examine the main difference between the antibound state
and the bound state discussed above. The corresponding Jost function
zero is at
$k_{1}=i\alpha_{1}$, but now $\alpha_{1}$ is negative.
The nonrelativistic relation
(\ref{lda_bound}) leads to a weaker coupling than in the previous case
($-1<\Lambda_{11}<0$). In Fig.~1b we show the energy dependence
of the phase shifts calculated in the relativistic and
nonrelativistic cases for two values of the range parameters
\mbox{$\beta_{1}$} as in Fig.~1a.
The phase shifts increase at the \mbox{$K\overline{K}$} threshold
starting from the zero
value, which means that the scattering length is positive.
Again we notice that the
relativistic effects are mainly important for the large values of the
parameter \mbox{$\beta_{1}$} and for higher energies. The phase shifts
calculated
relativistically are larger that the corresponding nonrelativistic
values.

   A special case in the \mbox{$K\overline{K}$} channel is the
existence of a resonance at the
complex energy $M_{S}-i\Gamma_{S}/2$, where $M_{S}$ is the resonance
energy
and $\Gamma_{S}$ is its width. In the complex momentum plane there are
two $S$--matrix poles in the lower half plane at $k_{1}=\pm
k_{R}-ik_{I}$, where $k_{I}>0$. They coincide with zeros of the Jost
function:
\begin{equation}
D_{K}(\pm k_{R}-ik_{I})=0.  \label{d_p_m}
\end{equation}

For our nonrelativistic and simple choice of the
\mbox{$K\overline{K}$} interaction
(Eq.~\ref{k_pot}) the interaction strength and the range parameter
are fixed,
since $\beta_{1}=k_{I}$ and $\Lambda_{11}=k_{R}^{2}/k_{I}^{2}>0$
(repulsive interaction). In the relativistic case, however, there
are two possibilities of choosing the parameters \mbox{$\beta_{1}$}
and \mbox{$\Lambda_{11}$} at a given resonance position.
One set of parameters is very close to the nonrelativistic set.
This is related
to the values of $k_{I}$ which in general must be smaller than about
100 MeV if
we would try to attribute the discussed resonance to the observed
rather narrow \mbox{$f_{0} (975)$} meson strongly coupled to the
\mbox{$K\overline{K}$} channel.
The second relativistic
solution is obtained for very large \mbox{$\beta_{1}$} value and
$\Lambda_{11}<0$ (attractive interaction).
The phase shifts are plotted in Fig.~1c and
their behaviour is very different in the two cases.
For smaller \mbox{$\beta_{1}$} values close to
70 MeV (upper curve) we at first observe an energy decrease and then
an increase of the $\delta_{K\overline{K}}$ function.
This means that the resulting scattering
lengths are negative. The phase shifts calculated for the relativistic
and
nonrelativistic propagators are very  similar.
They differ by less than one degree so the corresponding lines in
Fig.~1c are indistinguishable. The lower curve corresponds to the
second relativistic solution and the phase shifts increase staring
from $0^{\circ}$ at the threshold. In this case the scattering length
is positive.

   At the end of this section we return to the discussion of the bound
state,
especially its wave function in the momentum and configuration spaces.
An important parameter describing the bound \mbox{$K\overline{K}$}
system near the threshold is the root mean square radius parameter.
It has been discussed in
the nonrelativistic model \cite{cann89}, and its preferable value was
given as $\langle r_{S} \rangle^{1/2}=0.76$ fm.
The nonrelativistic form of the radial wave
function is simple and can be found in Ref.~\cite{cann89}. Its Fourier
transform $\Psi_{N}(p)$ describes the momentum distribution of the
\mbox{$K\overline{K}$} relative motion (see Ref.~\cite{eisenberg})
\begin{equation}
\Psi_{N}(p)=c_{N}\, G_{N}(p) g_{1}(p), \label{psi_n}
\end{equation}
where $G_{N}(p)=-m_{K}/(\alpha_{1}^{2}+p^{2})$ is the nonrelativistic
propagator
corresponding to the binding energy $E_{B}=\alpha_{1}^{2}/m_{K}$ and
$c_{N}$ is the normalization constant
\begin{equation}
c_{N}=-\frac{1}{2}\sqrt{\frac{\alpha_{1}\beta_{1}}{m_{K}}}\,
(\alpha_{1}+\beta_{1})^{3/2} \label{c_n}
\end{equation}
such that $\int{d^{3}k \mid \Psi_{N}(k) \mid^{2}}=1$.
Similarly in the relativistic case we write
\begin{equation}
\Psi_{R}(p)=c_{R}\, G_{R}(p) g_{1}(p), \label{psi_r}
\end{equation}
where $G_{R}(p)=1/(E-2\sqrt{p^{2}+m_{K}^{2}})$ is the relativistic
propagator
corresponding to the total energy $E=2\sqrt{m_{K}^{2}-\alpha_{1}^{2}}$
and $c_{R}$ is the normalization constant.
The relativistic wave function in the
configuration space $\Phi_{R}(r)$ is related to $\Psi_{R}(p)$ by the
Fourier transform. An important parameter describing the radial
extension of the \mbox{$K\overline{K}$} system is the root mean square
diameter
\begin{equation}
\langle r^{2} \rangle =\int{d^{3}r\,r^{2}\, \mid \Phi_{R}(r) \mid^{2}}
\label{diam}
\end{equation}
or the root mean square radius $\langle r_{S}^{2} \rangle ^{1/2}=
\frac{1}{2} \langle r^{2} \rangle ^{1/2}$.
The results of numerical calculations are given
in Table~\ref{tab_rms}. The first set of parameters
$\alpha_{1}$ and \mbox{$\beta_{1}$} is taken from Ref.~\cite{cann89}.
The second and third line correspond to the results of
fits described further in Sect. V. The root mean square radii are
small,
typically about 0.7 fm, so the \mbox{$K\overline{K}$} molecule has a
rather compact structure. This fact can have important consequences.
For example, in Ref.~\cite{close92} a value 1.2 fm
has been used to predict the width of the radiative \mbox{$\phi
\rightarrow
\mbox{$f_{0} (975)$} \gamma$} decay. If instead of 1.2 fm we use the
value 0.7 fm the corresponding width increases by a factor of 2.
Let us mention that the annihilation probability of the
\mbox{$K\overline{K}$} molecule into two
photons will be substantially increased for small
\mbox{$K\overline{K}$} radii (compare \cite{barnes}).
In Table II we can also notice a general trend: the
root mean square radii calculated for the relativistic propagator are
smaller by about 0.1
fm in comparison  with corresponding values for the nonrelativistic
case.
The difference comes from the behaviour of the radial wave functions
for the values of
$r$ smaller than about 0.7 fm, where the relativistic wave function
takes higher values than the nonrelativistic one.

\section{\mbox{$\pi\pi$} and \mbox{$K\overline{K}$} coupled channel
analysis}
\hspace{.5cm}
In this section we describe the coupled channel analysis performed for
the interacting \mbox{$\pi\pi$} and \mbox{$K\overline{K}$} pairs.
Starting from the \mbox{$\pi\pi$} threshold
we put the
upper energy limit at 1400 MeV, where one can still neglect the
coupling to other
channels with higher thresholds (as discussed in \cite{cann92}).
The interactions
(\ref{k_pot}--\ref{v12_pot}) in both channels have the forms as simple
as possible.
As it was mentioned in Sect.~III the rank--two potential allows us to
describe
a substantial increase of the \mbox{$\pi\pi$} phase shifts below 1 GeV
and the existence of the known \mbox{$f_{0} (1400)$} meson.
All the potential parameters have been
obtained by fitting the calculated phase shifts
\mbox{$\delta_{\mbox{$\pi\pi$}}$}, \mbox{$\delta_{K\overline{K}}$}\
and the inelasticity
parameter \mbox{$\eta$} to the experimental data.
Above the \mbox{$K\overline{K}$} threshold a
sum $\mbox{$\varphi$}=\mbox{$\delta_{\mbox{$\pi\pi$}}$}+
\mbox{$\delta_{K\overline{K}}$}$
is often used and a quantity $x=(1-\eta^{2})/4$ is introduced to
represent the inelasticity.
Except for the \mbox{$\pi\pi$} threshold data of
Refs.~\cite{rosselet},
\cite{belkov} we have used the data of Refs.~\cite{srinivas},
\cite{grayer} covering a wide energy band. The total number of the
\mbox{$\delta_{\mbox{$\pi\pi$}}$}\ points is 56. In addition, 17
points of the \mbox{$\eta$} dependence have
been taken from Ref.~\cite{cohen}. For the \mbox{$\varphi$}
dependence we use two distinct data
sets \cite{cohen}, \cite{estabrooks} which essentially differ in their
behaviour near the \mbox{$K\overline{K}$} threshold.
In the data \cite{pawlicki} used in the analysis of
Ref.~\cite{estabrooks} a constant increase of \mbox{$\varphi$} is
seen while in the data
of Ref.~\cite{cohen} some threshold decrease with energy can be
observed. As it
was shown in Sect. IV, this difference could be crucial for
understanding the
nature of \mbox{$f_{0} (975)$}, so we have performed two separate
fits to the collection of the
\mbox{$\delta_{\mbox{$\pi\pi$}}$},
\mbox{$\delta_{K\overline{K}}$}\ and \mbox{$\eta$} points.
The data set containing 16 \mbox{$\varphi$}--points of
Ref.~\cite{cohen} is further called set~1 while in the set~2 we
include 17 \mbox{$\varphi$}--numbers read from  Ref.~\cite{estabrooks}
(the total number of the data
points is 89 and 90 respectively for these two data sets).

   The fitted parameters are shown in Table III. The asymmetric errors
$\Delta_{+}$ and $\Delta_{-}$ correspond to an increase of the total
$\chi^{2}$ value by one unit.
Let us notice that all the channel coupling constants
\mbox{$\Lambda_{ii}$} are negative, which means that the interactions
are attractive in both channels.
The couplings between channels are small. The ranges of the form
factors are short since the \mbox{$\beta_{i}$} values are in the
GeV region. During the
fitting procedure we have found that it was convenient to use as an
independent
variable the product \mbox{$\Lambda_{33}$}\mbox{$\beta_{3}$} instead
of \mbox{$\beta_{3}$}. For both data sets this
product is very close to $-\pi m_{\pi} \simeq -431$ MeV.
This fact is not accidental as it is explained in Appendix.
The very large value of \mbox{$\beta_{3}$} is
compensated by very small values of the corresponding coupling
constants \mbox{$\Lambda_{33}$} and \mbox{$\Lambda_{13}$}.
The starting parameters \mbox{$\Lambda_{22}$} and
\mbox{$\beta_{2}$} in the
minimization procedure have been given in Sect.~III. Other coupling
constants were initially put equal to zero and \mbox{$\beta_{i}$}
parameters had the value 1 GeV.
In Table~\ref{tab_chi} we show $\chi^{2}$ values of two fits for
separated data parts. The data on \mbox{$\delta_{\mbox{$\pi\pi$}}$}\
and \mbox{$\eta$} are very well described by
both fits but the data on \mbox{$\varphi$} only by the fit to the
data set~1.

   In figures 2--4 the results of our fits to the data sets 1 and 2
are shown.
Theoretical \mbox{$\delta_{\mbox{$\pi\pi$}}$}\ and \mbox{$\eta$}
values  differ only slightly for both fits.
In Fig.~3 we notice an initial decrease of the \mbox{$\varphi$}
curves at the \mbox{$K\overline{K}$}
threshold,
so the data set~1 is clearly favoured by the model (compare also the
corresponding
$\chi^{2}$ values of Table~\ref{tab_chi}). Due to a very similar
monotonic increase of the \mbox{$\delta_{\mbox{$\pi\pi$}}$}\ phase
shifts above 1 GeV, the differences
in the \mbox{$\varphi$} behaviour for the data sets 1 and 2 are
related to various possible
trends of the $\delta_{K\overline{K}}$. As discussed in Sect.~IV the
trend seen
in the \mbox{$\varphi$}--data of Ref.~\cite{cohen} is connected with
a presence of the
quasibound \mbox{$K\overline{K}$} state but in order to exclude other
possibilities new
precise \mbox{$\delta_{K\overline{K}}$}\ measurements near the
threshold are needed.

   We have investigated a pole structure of the $S$ matrix elements
and the results are shown in Table~\ref{tab_res}.
A very wide \mbox{$f_{0} (500)$} meson (see Sect. III)
causes a systematic increase of the \mbox{$\delta_{\mbox{$\pi\pi$}}$}
phase shifts starting from the \mbox{$\pi\pi$}
threshold. The presence of the narrow \mbox{$f_{0} (975)$} manifests
itself by a
strong jump of the \mbox{$\delta_{\mbox{$\pi\pi$}}$}\ and the
enhancement of the inelasticity
function $x$ near the \mbox{$K\overline{K}$} threshold
(see Figs.~2 and 4). A further increase
of the \mbox{$\delta_{\mbox{$\pi\pi$}}$}\ and \mbox{$\varphi$} above
1.2 GeV is related to the presence of the
\mbox{$f_{0} (1400)$} meson.
Also the structure of the inelasticity around the energy 1.3 GeV is
caused by this state.
We should point out that the values of the resonance parameters may be
changed if a fit to another data sets is performed. We expect that
changes may
be much larger than the error ranges quoted in Table~\ref{tab_res}.
This is due to the fact that some experiments on the \mbox{$\pi\pi$}
and \mbox{$K\overline{K}$} scattering have
supplied contradictory results, as was illustrated by our choice of
the sets 1
and 2. We think, however, that the differences between data sets are
representative.
In particular the resonance parameters are not very much influenced
if we
compare results of two fits in Table~\ref{tab_res}.
Let us mention here that a moderate width of the \mbox{$f_{0} (1400)$}
is obtained together
with the presence of the wide \mbox{$f_{0} (500)$} meson.
The calculated masses and widths of the
\mbox{$f_{0} (1400)$} meson are in very good agreement with the
corresponding values obtained
in ref. \cite{cohen} ($M=1425 \pm 15$ MeV and $\Gamma=160
\pm 30$ MeV). In the absence of the former
state the systematic increase of the \mbox{$\pi\pi$} phase shifts
below the
energy of 1 GeV should be caused by a very wide \mbox{$f_{0} (1400)$}
state (see for
example Ref.~\cite{cann89}) or by another wide state at the energy of
about 1 GeV.

   In Fig.~5 the $S_{\mbox{$\pi\pi$}}$ matrix element singularities
are presented
using the complex variable $z$ (compare \cite{kato} and \cite{cann89})
\begin{equation}
z=\frac{k_{1}+k_{2}}{\sqrt{m_{K}^{2}-m_{\mbox{$\pi$}}^{2}}}.
\label{z_def}
\end{equation}
The positions of poles and zeros have been calculated from the fit to
the data
set~1. A very similar structure of the $S_{\mbox{$\pi\pi$}}$ has been
found for the fit to the data set~2.
The energy sheets on the $z$--plane are
defined by the signs of the imaginary parts of the $k_{2}$ and $k_{1}$
momenta $(Im\,k_{2}, Im\,k_{1})$ as follows:
\mbox{I (+,+)}, \mbox{II (--,+)}, \mbox{III (--,--)},
\mbox{IV (+,--)}.
In Fig.~5 we notice two poles on sheet II labelled as 1 and 2. They
correspond to the \mbox{$f_{0} (500)$} and \mbox{$f_{0} (975)$}
resonances. The \mbox{$f_{0} (1400)$} resonance
position on sheet III is indicated by 3.
Those poles are the nearest singularities lying
close to the physical region. The \mbox{$\pi\pi$} threshold region
is also
strongly influenced by the cuts located on the imaginary $z$-axes.
In the momentum space
these cuts are on the $k_{i}$ axes (from $\pm \,im_{i}$ to $\pm
\,i\infty$). Their
origin is due to the presence of the logarithmic and square root
functions in the
\mbox{$J_{ij}$} integrals (see Appendix). In the nonrelativistic
case these integrals
have much simpler structure and there are no cuts in the $z$--plane.
Two poles
at Re$\,z>0$ are particularly close to the $\mid\! z\! \mid = 1$
circle: pole 2
associated with the \mbox{$f_{0} (975)$} resonance and the another
pole lying on sheet III with a zero almost superposed on its top.
Pole 2 can be related to the
antibound state discussed in Sect.~IV. If the coupling between
channels is
switched off then the poles move to the $\mid\!\! z\!\! \mid =1$
circle. The positions of
other zeros or poles are connected with the discussed
\mbox{$f_{0} (500)$} and \mbox{$f_{0} (1400)$}
resonances. In particular the structure of the inelasticity function
shown in
Fig.~4 is a result of an interplay of the $S$--matrix zeros and poles.
The $S_{\mbox{$\pi\pi$}}$ matrix element also has two
second order zeros related to the form of the \mbox{$\pi\pi$}
potential. They correspond
to the values of \mbox{$k_{2}=-i\beta_{2}$} and \mbox{$k_{2}=
-i\beta_{3}$} and due to large values of \mbox{$\beta_{2}$}
and \mbox{$\beta_{3}$} two of them lay very close to the origin of the
$z$--plane and the other two very far from that point.
These poles are not shown in Fig.~5.

   We have also performed fits with only one term in the
\mbox{$\pi\pi$}
potential i.e. with $\mbox{$\lambda_{33}$}=0$ in Eq.~(\ref{p_pot}).
The $\chi^{2}$ fit was poor.
In this case the $S_{\mbox{$\pi\pi$}}$ matrix element has no pole
corresponding to the \mbox{$f_{0} (1400)$} meson and the
\mbox{$\delta_{\mbox{$\pi\pi$}}$}\ and \mbox{$\varphi$} phase
shifts have a very flat behaviour above the energy 1.2 GeV.

   Similar fits as for the relativistic form have been done using
a nonrelativistic form of the Jost function.
The coupling constants \mbox{$\Lambda_{22}$} and \mbox{$\Lambda_{33}$}
had then opposite signs and the \mbox{$f_{0} (975)$} meson also
appeared as a \mbox{$K\overline{K}$} bound state.
We have found that apart from the \mbox{$f_{0} (975)$} meson there
was only one very
wide ($\Gamma\approx 1300$ MeV) state at the energy of about 850 MeV.
The $\chi^{2}$ value for the fit was 283 for 89 points so it was much
worse than in the relativistic case. Energy dependence of the
\mbox{$\delta_{\mbox{$\pi\pi$}}$}\ , \mbox{$\varphi$} and
$x$ was similar to that one in the relativistic fit with only one term
in the
\mbox{$\pi\pi$} potential. Like in Ref.~\cite{cann89}, the $S$ matrix
also has two poles on the negative part of the imaginary axis $k_{2}$
but no poles corresponding to the \mbox{$f_{0} (500)$}.

   Scattering lengths $a_{\mbox{$\pi\pi$}}$ and $a_{K\overline{K}}$
are very important
quantities describing the near threshold \mbox{$\pi\pi$} and
\mbox{$K\overline{K}$} interactions.
Comparison of the values obtained by us with experimental measurements
and other
theoretical estimations is shown in Table~\ref{tab_scatt}.
Good agreement of the predicted $a_{\mbox{$\pi\pi$}}$ values for the
relativistic case with other data is found.
In the nonrelativistic case there are
two $S$--matrix poles on the $k_{2}$ imaginary axis and one of them is
relatively
close to the \mbox{$\pi\pi$} threshold, which gives a large positive
contribution to the scattering length.
Therefore the scattering length is much larger than in the
relativistic case.
An appearance of the imaginary part of the \mbox{$K\overline{K}$}
scattering length is related to
the fact that above the \mbox{$K\overline{K}$} threshold the
annihilation process into two pions is possible.
The value of this imaginary part is evaluated using a low momentum
approximation of the inelasticity \mbox{$\eta$}
\begin{equation}
\eta \approx 1-2 b k_{1}+O(k_{1}^{2}), \label{eta_approx}
\end{equation}
where the parameter $b$ is equal to Im$a_{K\overline{K}}$ (see also
Ref.~\cite{cann92}). The negative value of the real part of the
\mbox{$K\overline{K}$} scattering
length is caused by an influence of a single pole lying close to the
\mbox{$K\overline{K}$} threshold on the sheet II (see Fig.~5).
This pole is related with the narrow
\mbox{$f_{0} (975)$} meson. We expect that in a case when two poles
are close to the
\mbox{$K\overline{K}$} threshold on sheets II and III
(as discussed in Ref.~\cite{ral92}) the absolute value of
the real part of the \mbox{$K\overline{K}$} scattering length can be
much smaller. The reason is
that the contributions of such poles have opposite signs so they
cancel to a
large extend. The cancellation is complete if the poles lie
symmetrically in the
kaon momentum space (at $k_{1}$ and --$k_{1}$).
The precise experimental measurements
of this quantity could distinguish between these two possibilities.
There is one experimental estimation of the \mbox{$K\overline{K}$}
scattering length based on the data of
ref. \cite{wetzel} which gives numbers similar to our findings.
Unfortunately,
the \mbox{$K\overline{K}$} phase shifts used by Wetzel et al. do not
satisfy a very general
requirement $\mbox{$\delta_{K\overline{K}}$}(-k_{1})=
-\mbox{$\delta_{K\overline{K}}$}(k_{1})$, so the values of the
scattering lengths
obtained in \cite{wetzel} could be questioned. The $K^{+}K^{-}$
scattering
lengths quoted in ref. \cite{dumbrajs} and given in Table
\ref{tab_scatt} are
theoretical estimations used in the calculations of the properties
specific to the kaonic atoms.
Contrary to the \mbox{$K\overline{K}$} case, more information on the
\mbox{$\pi\pi$}
scattering length is available. Our values for the relativistic fits
are in
good agreement with other determinations (see Table \ref{tab_scatt}).

   Due to the channel coupling the \mbox{$f_{0} (975)$} and
\mbox{$f_{0} (1400)$} mesons can decay
into the \mbox{$\pi\pi$} and \mbox{$K\overline{K}$} pairs.
The \mbox{$f_{0} (975)$} meson being presumably a
\mbox{$K\overline{K}$} bound state becomes
unstable due to the annihilation process \mbox{$\mbox{$K\overline{K}$}
\rightarrow\mbox{$\pi\pi$}$}.
The branching ratio of its decay into the \mbox{$K\overline{K}$}
channel is
\begin{equation}
B=\frac{\Gamma_{K\overline{K}}}{\Gamma_{tot}}, \label{br_ratio}
\end{equation}
where $\Gamma_{tot}$ denotes the total and $\Gamma_{K\overline{K}}$
the \mbox{$K\overline{K}$} fractional decay width.
If we wish to compare the experimental \mbox{$f_{0} (975)$} branching
ratio with the theoretical one, we must take into account the fact
that due to the
proximity of the \mbox{$K\overline{K}$} threshold the
\mbox{$f_{0} (975)$} resonance has no a typical
Breit-Wigner shape.
As it was discussed in Ref.~\cite{cann92} the average branching ratio
$B_{av}$ of the \mbox{$f_{0} (975)$} reads
\begin{equation}
B_{av}=\frac{1}{2(M_{max}-M_{s})} \int_{M_{th}}^{M_{max}} dE
(1-\eta^{2}),
\label{b_av}
\end{equation}
where $M_{max}$, $M_{s}$ and $M_{th}$ denote respectively the upper
limit of
integration, mass of the meson and mass of the \mbox{$K\overline{K}$}
threshold. In the numerical
calculations we put $M_{max}=4\Gamma_{tot}$.
We obtain $B_{av}=18.0\pm 0.7\%$
for the fit to the data set~1 and $B_{av}=18.4\pm 0.6\%$ for the
fit to the data set~2
in agreement with the Particle Data Group value $21.9\pm 2.4\%$
\cite{pdg92}.
In the case of the \mbox{$f_{0} (1400)$} meson we have performed
the integration from
$M_{s}-\Gamma_{tot}$ to $M_{s}+\Gamma_{tot}$ substituting
$2\Gamma_{tot}$ in
place of $2(M_{max}-M_{s})$ in Eq.~(\ref{b_av}).
For the \mbox{$f_{0} (1400)$}$\rightarrow$\mbox{$K\overline{K}$}
branching ratio we obtain 16$\pm$1\% for both data sets in
agreement with \cite{cohen}.
Let us notice that the \mbox{$f_{0} (1400)$} resonance mass is
slightly higher than the upper energy limit 1.4 GeV used in the fits.
This fact may lead to some uncertainties
in the predicted values of the \mbox{$f_{0} (1400)$} branching ratios.
The Particle Data Group value is $6.4_{-2.5}^{+1.9}\%$ \cite{pdg92}.
It is
entirely based on the analysis done in Ref.~\cite{gorlich} in the
energy range
from 1100 to 1420 MeV. The corresponding value of the branching ratio
calculated
for the parameter set 1 in the same energy range is as high as 30.3\%.
This
is a result of the existing discrepancy between the $s$-wave
intensities
obtained by the Argonne group \cite{cohen} and the CERN-Cracow-Munich
group \cite{gorlich}.
In our analysis we have chosen the Argonne data, since they
include points closer to the \mbox{$K\overline{K}$} threshold.
For energies higher than 1.4 GeV an influence of other
scalar--isoscalar states
such as $f_{0}(1525)$, $f_{0}(1590)$ or even $f_{0}(1710)$
(see references in \cite{pdg92}) may be important.
We have also tried to include in our
analysis the energy range from 1.4 GeV to 1.5 GeV and we could not
simultaneously fit the \mbox{$\delta_{\mbox{$\pi\pi$}}$}\ and
\mbox{$\varphi$} phase shifts.
The \mbox{$\delta_{\mbox{$\pi\pi$}}$}\ data in this energy
region favour wider \mbox{$f_{0} (1400)$} meson; the sum
$\mbox{$\varphi$}=\mbox{$\delta_{\mbox{$\pi\pi$}}$}+
\mbox{$\delta_{K\overline{K}}$}$ which increases
steeply with energy requires an existence of a narrow state.

\section{Conclusions}
\hspace{.5cm}
We have performed the analysis of the isoscalar spin zero
\mbox{$\pi\pi$} and
\mbox{$K\overline{K}$} coupled channel interactions.
The model was based on a separable potential formalism.
Two channel scattering amplitudes have been evaluated from the
coupled equations of the Lippmann--Schwinger type. Using the
relativistic
propagators we have obtained very good fits to the experimental
\mbox{$\pi\pi$} and \mbox{$K\overline{K}$} scattering data.
It has been found that the potentials in the \mbox{$\pi\pi$} and
\mbox{$K\overline{K}$} channels are attractive and have short range
form factors. The $\chi^{2}$
fits in the relativistic case were 3 or 4 times better than in
the nonrelativistic case.
In contrast to the nonrelativistic calculations (see \cite{cann89}
and Sect.~V) new \mbox{$f_{0} (500)$} state has been found in the
\mbox{$\pi\pi$}
interactions treated relativistically. Its large width about 500 MeV
is in agreement with other estimations.
In the relativistic version of our model the mass and
width of the \mbox{$f_{0} (1400)$} meson are well described
(see Table \ref{tab_res}).
Relativistic effects also play the important role near the
\mbox{$\pi\pi$} threshold.
The \mbox{$\pi\pi$} scattering length is significantly smaller
than in the nonrelativistic case and is in agreement with
the experimental data and other models (see Table \ref{tab_scatt}).

   We have also been interested in describing the
\mbox{$K\overline{K}$} interaction.
Two different data sets on the \mbox{$K\overline{K}$} phase shifts
have been taken into account.
Our solutions indicate the existence of a quasibound state below the
\mbox{$K\overline{K}$} threshold which we identify with the
\mbox{$f_{0} (975)$} meson.
Thus this state does not seem to be a typical $q\overline{q}$ meson.
Some observables, e.g. the \mbox{$K\overline{K}$} scattering length
and
root mean square radius of the \mbox{$f_{0} (975)$} have been
evaluated. We predict the large
negative value of the \mbox{$K\overline{K}$} scattering length about
$-1.7$ fm and
the corresponding imaginary part of the order of 0.6 fm (Table VI).
In our calculations the \mbox{$f_{0} (975)$} state appears as the
\underline{compact} \mbox{$K\overline{K}$} system with the root mean
square radius of
about 0.7 fm, so it is not a \mbox{$K\overline{K}$} molecule of the
deuteron size.
Analysis of the \mbox{$K\overline{K}$} channel leads to a conclusion
that at the \mbox{$K\overline{K}$} threshold
the relativistic corrections to the \mbox{$K\overline{K}$} scattering
length or the root
mean square radius are of the order of 10\% and they gradually grow
with increasing energy.

   The existing data above the \mbox{$K\overline{K}$} threshold are
still controversial.
Comparison of our expectations with new experimental data may be done
in near
future and a new light on the nature of the \mbox{$f_{0} (975)$} and
$a_{0}(980)$ mesons can
be shed on if very precise measurements near the
\mbox{$K\overline{K}$} production threshold like
those planned at COSY \cite{oelert} are performed.
Very good energy resolution would enable us to take into account a
mass splitting
between the charged $K^{-}K^{+}$ and neutral $K^{0}\overline{K^{0}}$
modes.
Due to different interactions of the $K^{-}$ and $K^{+}$ with nucleons
in the GeV
region the molecular picture of the \mbox{$f_{0} (975)$} can also be
verified by studying its
interaction inside nuclear matter (see also Ref.~\cite{cann92}).

\vspace{.5cm}
   Acknowledgments

   This work has been partially supported by Polish Committee for
Scientific Research (grant No 2 0198 9101). Discussions with L.
G\"orlich, A. D. Martin, M. R\'o\.za\'nska, K. Rybicki and J. Turnau
are gratefully acknowledged. We thank very much D. Morgan for his
computer communications.

\appendix
\renewcommand{\theequation}{A\arabic{equation}}
\renewcommand{\thesection}{}

\section{Appendix}
\setcounter{equation}{0}
\hspace{.5cm}
In this appendix we give explicit expressions for the $t$--matrix
elements (\ref{t})
in terms of the coupling constants \mbox{$\lambda_{ij}$} and the
integrals \mbox{$I_{ij}$}
(Eqs. (\ref{I_ii}) and (\ref{I_kl})):
\begin{eqnarray}
\hspace{-1.0cm}t_{11}&=&D^{-1}[\lambda_{11}-(\lambda_{11}\lambda_{22}-
\lambda_{12}^{2})I_{22}-(\lambda_{11}\lambda_{33}-
\lambda_{13}^{2})I_{33}+2\lambda_{12}\lambda_{13}I_{23}\nonumber\\
 &+& (\lambda_{11}\lambda_{22}\lambda_{33}-
\lambda_{33}\lambda_{12}^{2}-
\lambda_{22}\lambda_{13}^{2})(I_{22}I_{33}-I_{23}^{2})\,],
\label{at11}\\
\hspace{-1.0cm}t_{12} & = & D^{-1}(\lambda_{12}-\lambda_{12}
\lambda_{33}I_{33}+\lambda_{22}\lambda_{13}I_{23}),\label{at12}\\
\hspace{-1.0cm}t_{13} & = & D^{-1}(\lambda_{13}-\lambda_{13}
\lambda_{22}I_{22}+\lambda_{33}\lambda_{12}I_{23}),\label{at13}\\
\hspace{-1.0cm}t_{22}&=&D^{-1}\{\lambda_{12}I_{11}[\lambda_{12}
(1-\lambda_{33}I_{33})+\lambda_{22}\lambda_{13}I_{23}]\nonumber\\
&+ & \lambda_{22}(1-\lambda_{11}I_{11})(1-\lambda_{33}I_{33})
-\lambda_{22}\lambda_{13}I_{11}(\lambda_{12}I_{23}+
\lambda_{13}I_{33})\},
\label{at22}\\
\hspace{-1.0cm}t_{23}&=&D^{-1}[\lambda_{22}\lambda_{33}I_{23}+
\lambda_{12}
\lambda_{13}I_{11}\nonumber\\
& & -(\lambda_{11}\lambda_{22}
\lambda_{33}-\lambda_{22}\lambda_{13}^{2}-
\lambda_{33}\lambda_{12}^{2})I_{11}I_{23}],\label{at23}\\
\hspace{-1.0cm}t_{33}&=&D^{-1}\{\lambda_{13}I_{11}[\lambda_{13}
(1-\lambda_{22}I_{22})+\lambda_{33}\lambda_{12}I_{23}]
\nonumber\\
&+ & \lambda_{33}(1-\lambda_{11}I_{11})(1-\lambda_{22}I_{22})
\mbox{}-\lambda_{33}\lambda_{12}I_{11}(\lambda_{13}I_{23}+
\lambda_{12}I_{22})\}\label{at33}.
\end{eqnarray}
In Eqs.~(\ref{at11})--(\ref{at33}) $D$ denotes the function defined
in Eq.~(\ref{DKPC}).
The integrals (~\ref{I_ii}) and (~\ref{I_kl}) have also been used in
\cite{siegel88} to calculate the \mbox{$K^{-}$--$N$} potential but
analytical
formulas have been given only when the propagator $G$ in
Eq.~(\ref{G_def}) was
nonrelativistic (see also \cite{alberg76}). We show below that in the
relativistic case the analytical calculations can be performed as
well. A very helpful substitution is:
\mbox{$y=\sqrt{s^{2}/(s^{2}+m_{i}^{2})}$}.
Then after some straightforward algebraic operations the integrals
\mbox{$I_{ij}$} can be rewritten in the terms of the integral
\begin{equation}
F(x^{2})=\int_{0}^{1}\frac{dy}{1-y^{2}x^{2}+i\mbox{$\epsilon$}}.
\label{aF_def}
\end{equation}
If $x^{2}$ is real then
\begin{equation}
F(x^{2})=\left\{ \begin{array}{llll}
         \frac{\mbox{$1$}}{\mbox{$2x$}}\,\ln\left(\frac{\mbox{$x$}+
\mbox{$1$}}{\mbox{$x$}-\mbox{$1$}}\right)-
\frac{\mbox{$i\pi$}}{\mbox{$2x$}}
                                         & \mbox{if $x^{2}>1$},
\vspace{.2cm} \\
\vspace{.2cm}
         \frac{\mbox{$1$}}{\mbox{$2x$}}\,
\ln\left(\frac{\mbox{$1$}+\mbox{$x$}}{\mbox{$1$}-\mbox{$x$}}\right)
                                         & \mbox{if $0<x^{2}<1$}, \\
\vspace{.2cm}
		1                        & \mbox{if $x^{2}=0$}, \\
\vspace{.2cm}
         \arctan(\mid x \mid)/\mid x \mid
		                         & \mbox{if $x^{2}<0$}.
		\end{array}
	\right.
\label{aF_res}
\end{equation}
Finally the integrals \mbox{$I_{ij}$} can be expressed in terms of
\mbox{$J_{ij}$} (Eq.~(\ref{J_def})):
\begin{eqnarray}
\lefteqn{J_{ii}(k_{i})=-\frac{E_{i}}{4m_{i}(1-ia_{i})^{2}}\,-\,
\frac{2}{\pi b_{i}(1+a_{i}^{2})^{2}}} \nonumber\samepage \\
&&\times\left\{\frac{1}{2}\,(1+a_{i}^{2})+\left[-\frac{1}{2}\,
(1+a_{i}^{2})\,
b_{i}^{2}+a_{i}^{2}C_{i}^{2}\right]d_{i}-a_{i}^{2}C_{i}^{2}H_{i}
\right\},
\label{aI_ii}
\end{eqnarray}
\begin{eqnarray}
\lefteqn{\hspace{-1.3cm}J_{23}(k_{2})=
-\frac{E_{2}}{2m_{2}(1-ia_{2})(1-ia_{3})}\,
\frac{\sqrt{\beta_{2}\beta_{3}}}
{\beta_{2}+\beta_{3}}+\frac{2}{\pi\sqrt{b_{2}b_{3}}(1+a_{2}^{2})
(1+a_{3}^{2})}
} \nonumber \\
&&\hspace{-1.5cm}\times\left\{\frac{E_{2}^{2}}{4\beta_{2}
\beta_{3}}H_{2}+
\frac{\beta_{2}\beta_{3}}{\beta_{3}^{2}-\beta_{2}^{2}}
\left[d_{3}(1+a_{2}^{2})(b_{3}^{2}-1)-d_{2}(1+a_{3}^{2})(b_{2}^{2}-1)
\right]\right\}.
\label{aI_23}
\end{eqnarray}
In (\ref{aI_ii}) and (\ref{aI_23})
\begin{equation}
E_{i}=2\sqrt{k_{i}^{2}+m_{i}^{2}},\;C_{i}=\frac{E_{i}}{2k_{i}},
\;H_{i}\equiv F(C_{i}^{2}) \label{aech}
\end{equation}
and
\begin{equation}
a_{i}=\frac{k_{i}}{\beta_{i}},\;b_{i}=\frac{m_{i}}{\beta_{i}},
\;d_{i}\equiv F(1-b_{i}^{2}),\;\;i=1,2,3.
\label{aabdi}
\end{equation}
In (\ref{aech}) and (\ref{aabdi}) the $m_{i}$ are defined by
(\ref{mi_def}) and $k_{3}\equiv k_{2}$.

   The nonrelativistic limit of the integrals (\ref{I_ii}) and
(\ref{I_kl})
can be obtained from (\ref{aI_ii}) and (\ref{aI_23}) when
$m_{i}\longrightarrow\infty$. Physically it means that the
nonrelativistic limit
is achieved when both $k_{i}$ ($i=1,2$) and $\beta_{j}$ ($j=1,2,3$)
are much smaller than $m_{i}$. Then
\begin{equation}
J_{ii}(k_{i})=-\frac{1}{(1-ia_{i})^{2}}
\label{aI_ii_n}
\end{equation}
and
\begin{equation}
J_{23}(k_{2})=-\frac{2\sqrt{\beta_{2}\beta_{3}}}
{(\beta_{2}+\beta_{3})(1-ia_{2})(1-ia_{3})}.
\label{aI_23_n}
\end{equation}

   Let us discuss the limit $\beta_{3}\longrightarrow\infty$. For very
large \mbox{$\beta_{3}$} the integral $J_{33}$ is proportional to
\mbox{$\beta_{3}$}:
\begin{equation}
J_{33} \simeq - \frac{\beta_{3}}{\mbox{$\pi$} m_{\pi}}.
\label{b_J_33}
\end{equation}
In order to compensate too large values of $J_{33}$ in
Eq.~(\ref{P}-\ref{d}) we
can multiply it by a sufficiently small \mbox{$\Lambda_{33}$} value
which should be inversely proportional to \mbox{$\beta_{3}$}.
If $\beta_{3}\longrightarrow\infty$ and
$\Lambda_{33}\beta_{3}=-\pi m_{\pi}$ then the term
$1-\Lambda_{33}J_{33}$ in
Eq.~(\ref{P}) does not grow with \mbox{$\beta_{3}$}.

   In the limit $\beta_{3}\longrightarrow\infty$ the integral
\mbox{$J_{23}$} vanishes in
Eq.~(\ref{d}) and the Jost function $D_{\mbox{$\pi$}}(k_{2})$ is
approximately
a product of two terms \mbox{$(1-\Lambda_{22}J_{22})
(1-\Lambda_{33}J_{33})$}.
The zero of the first term can be attributed to the
\mbox{$f_{0} (500)$} meson as discussed in Sect. III.
For the values \mbox{$\beta_{3}$} and \mbox{$\Lambda_{33}$} as given
in Table III the second term (together with the small
$J_{23}^{2}$ term) allows us to obtain
the second $D_{\mbox{$\pi$}}$ zero relatively close to the real axis
$k_{2}$.
In Table~\ref{tab_res} we assign this zero to the
\mbox{$f_{0} (1400)$} meson.

\clearpage

\renewcommand{\thetable}{\Roman{table}}
\begin{table}
\centering
\caption{\mbox{$K\overline{K}$} coupling constants
\mbox{$\Lambda_{11}$} and scattering lengths $a$
corresponding to the \mbox{$K\overline{K}$} bound state at 975 MeV
(without coupling to the \mbox{$\pi\pi$} channel).}
\vspace{.4cm}

\begin{tabular}{|r|r|r|r|r|c|}
\hline
\multicolumn{1}{|c|}{\mbox{$\beta_{1}$}} &
\multicolumn{2}{|c|}{nonrelativistic} &
\multicolumn{2}{|c|}{relativistic} &  \\
\cline{2-3}
\cline{4-5}
\multicolumn{1}{|c|}{(MeV)} &
\multicolumn{1}{|c|}{\mbox{$\Lambda_{11}$}} &
\multicolumn{1}{|c|}{$a_{N}$ (fm)} &
\multicolumn{1}{|c|}{\mbox{$\Lambda_{11}$}} &
\multicolumn{1}{|c|}{$a_{R}$ (fm)} &
\multicolumn{1}{|c|}{$\mid \frac{\mbox{$a_{R}$}}
{\mbox{$a_{N}$}}-1 \mid \cdot 100 \%$} \\
\hline
150 & -2.55 & -4.32 & -2.48 & -4.28 & 1 \\
500 & -1.39 & -2.81 & -1.18 & -2.70 & 4 \\
2000 & -1.09 & -2.35 & -0.54 & -2.05 & 13 \\
\hline
\end{tabular}
\label{tab_kk_sc_coupl}
\end{table}


\begin{table}
\centering
\caption{Root mean square radii of the \mbox{$K\overline{K}$}
wave function.}
\vspace{.4cm}

\begin{tabular}{|r|r|r|c|c|}
\hline
& & &
\multicolumn{1}{|c|}{nonrelativistic} &
\multicolumn{1}{|c|}{relativistic} \\
\multicolumn{1}{|c|}{$E$} &
\multicolumn{1}{|c|}{$\alpha_{1}$} &
\multicolumn{1}{|c|}{\mbox{$\beta_{1}$}} &
\multicolumn{1}{|c|}{$\langle r_{S}^{2} \rangle ^{1/2}$} &
\multicolumn{1}{|c|}{$\langle r_{S}^{2} \rangle ^{1/2}$} \\
\multicolumn{1}{|c|}{MeV} &
\multicolumn{1}{|c|}{MeV} &
\multicolumn{1}{|c|}{MeV} &
\multicolumn{1}{|c|}{fm} &
\multicolumn{1}{|c|}{fm} \\
\hline
971.65 & 98.90 & 2000 & 0.76 & 0.66 \\
973.36 & 94.51 & 1496 & 0.81 & 0.73 \\
973.71 & 93.59 & 2177 & 0.80 & 0.69 \\
\hline
\end{tabular}
\label{tab_rms}
\end{table}

\clearpage

\begin{table}
\centering
\caption{Model parameters and their up ($\Delta_{+}$) and down
($\Delta_{-}$) errors.}
\vspace{.4cm}

\begin{tabular}{|l|r|r|r|r|r|r|}
\hline
\multicolumn{1}{|c|}{fitted} &
\multicolumn{3}{|c|}{set 1} &
\multicolumn{3}{|c|}{set 2} \\
\cline{2-4}
\cline{5-7}
\multicolumn{1}{|c|}{parameters} &
\multicolumn{1}{|c|}{values} &
\multicolumn{1}{|c|}{$\Delta_{+}$} &
\multicolumn{1}{|c|}{$\Delta_{-}$} &
\multicolumn{1}{|c|}{values} &
\multicolumn{1}{|c|}{$\Delta_{+}$} &
\multicolumn{1}{|c|}{$\Delta_{-}$} \\
\hline
\mbox{$\Lambda_{11}$} & -0.658 & 0.030 & -0.031 & -0.511 & 0.075 &
-0.078 \\
\mbox{$\Lambda_{22}$} & -0.201 & 0.003 & -0.003 & -0.201 & 0.004 &
-0.004 \\
$\Lambda_{33} \times 10^{5}$ & -7.46 & 2.32 & -3.51 & -8.95 & 3.11 &
-5.55 \\
\mbox{$\Lambda_{12}$} & 0.0363 & 0.0021 & -0.0025 & 0.0251 & 0.0064 &
-0.0048 \\
$\Lambda_{13} \times 10^{6}$ & 3.0 & 2.5 & -1.0 & 2.8 & 2.9 &
-0.9 \\
\mbox{$\beta_{1}$} (GeV) & 1.496 & 0.115 & -0.082 & 2.177 & 0.528 &
-0.402 \\
\mbox{$\beta_{2}$} (GeV) & 1.162 & 0.052 & -0.051 & 1.141 & 0.052 &
-0.052 \\
\mbox{$\beta_{3}$}$\times$\mbox{$\Lambda_{33}$} (MeV) & -431.162 &
0.070 & -0.040
& -431.145 & 0.065 & -0.035 \\
\hline
$\mbox{$\beta_{3}$} \times 10^{-3}$ (GeV) & 5.8 & 2.6 & -1.9
& 4.8 & 2.6 & -1.8 \\
\hline
\end{tabular}
\label{tab_par}
\end{table}


\begin{table}
\centering
\caption{$\chi^{2}$ values for the fits to the data sets 1 and 2.}
\vspace{.4cm}

\begin{tabular}{|c|r|c|r|r|}
\hline
\multicolumn{1}{|c|}{Set} &
\multicolumn{1}{|c|}{$\chi^{2}$} &
\multicolumn{1}{|c|}{$\chi^{2}$} &
\multicolumn{1}{|c|}{$\chi^{2}$} &
\multicolumn{1}{|c|}{$\chi^{2}$} \\
\multicolumn{1}{|c|}{No}  &
\multicolumn{1}{|c|}{total} &
\multicolumn{1}{|c|}{$\delta_{\mbox{$\pi\pi$}}$ data} &
\multicolumn{1}{|c|}{$\varphi$ data} &
\multicolumn{1}{|c|}{$\eta$ data} \\
\hline
1 & 75.2 & 52.2 & 6.7 & 16.4 \\
2 & 100.8 & 50.5 & 34.2 & 16.1 \\
\hline
\end{tabular}
\label{tab_chi}
\end{table}

\clearpage

\begin{table}
\centering
\caption{Masses and widths of resonances obtained in fits to the
data sets 1
and 2 compared with values of the Particle Data Group [34]
and Ref.~[39] for \mbox{$f_{0} (500)$}. }
\vspace{.4cm}

\begin{tabular}{|l|r|r|r|r|r|r|}
\hline
&\multicolumn{2}{|c|}{set 1} &
 \multicolumn{2}{|c|}{set 2} &
 \multicolumn{2}{|c|}{Particle Data Group}   \\
\cline{2-3}
\cline{4-5}
\cline{6-7}
\multicolumn{1}{|c|}{pole} &
\multicolumn{1}{|c|}{M$\,$(MeV)} &
\multicolumn{1}{|c|}{$\Gamma\,$(MeV)} &
\multicolumn{1}{|c|}{M$\,$(MeV)} &
\multicolumn{1}{|c|}{$\Gamma\,$(MeV)} &
\multicolumn{1}{|c|}{M$\,$(MeV)} &
\multicolumn{1}{|c|}{$\Gamma\,$(MeV)} \\
\hline
\mbox{$f_{0} (500)$} & 506$\pm 10$ & 494$\pm 5$ & 505$\pm 10$ &
497$\pm 5$
& $\leq 700$ & $\geq 600$ \\
\mbox{$f_{0} (975)$}  & 973$\pm 2$  & 29$\pm 2$ & 974$\pm 2$ &
30$\pm 1$
& 974.1$\pm 2.5$ & 47$\pm 9$ \\
\mbox{$f_{0} (1400)$} & 1430$\pm 5$ & 145$\pm 25$ &
1428$^{+13}_{-7}$ &
157$^{+43}_{-29}$
& $\sim 1400$ & $150\div400$ \\
\hline
\end{tabular}
\label{tab_res}
\end{table}


\begin{table}
\centering
\caption{Comparison of the \mbox{$\pi\pi$} and \mbox{$K\overline{K}$}
scattering lengths
obtained in the
present work (sets 1, 2 and nonrelativistic fit) with other
determinations.}
\vspace{.4cm}

\begin{tabular}{|c|c|c|c|}
\hline
Source & $a_{\mbox{$\pi\pi$}}$ ($m_{\mbox{$\pi$}}^{-1}$) &
Re($a_{K\overline{K}}$) (fm) &
Im($a_{K\overline{K}}$) (fm) \\
\hline
set 1 & 0.172$\pm$0.008 & -1.73$\pm$0.07 & 0.59$\pm$0.04 \\
set 2 & 0.174$\pm$0.008 & -1.58$\pm$0.09 & 0.61$\pm$0.04 \\
nonrel. fit & 0.40 & -1.70 & 0.78 \\
\cite{dumbrajs} & --- & -1.15 & 1.80 \\
\cite{lowe} & 0.207$\pm 0.028$ & --- & --- \\
\hline
\end{tabular}
\label{tab_scatt}
\end{table}

\clearpage

\begin{center}
FIGURE CAPTIONS\\
\end{center}

\addvspace{2.54cm}

\begin{itemize}
\item Fig.~1a. \mbox{$K\overline{K}$} scattering phase shifts versus
energy in the
presence of the
bound state at 975 MeV. Solid lines correspond to the relativistic
\mbox{$K\overline{K}$}
propagator, while the dashed line to the nonrelativistic propagator.
Two upper
lines are calculated for \mbox{$\beta_{1}$}=2000 MeV, two lower
lines -- for \mbox{$\beta_{1}$}=150 MeV.
\item Fig.~1b Same as in Fig. 1a but for the antibound state at
975 MeV.
\item Fig.~1c \mbox{$K\overline{K}$} scattering phase shifts versus
energy in the presence of the
resonance at $M_{s}=993$ MeV of the width $\Gamma=46$ MeV.
Both curves are calculated for the relativistic propagator: the upper
one corresponds to \mbox{$\beta_{1}$}=72 MeV and
\mbox{$\Lambda_{11}$}=1.16 and the lower
one to \mbox{$\beta_{1}$}=10.47 GeV and \mbox{$\Lambda_{11}$}=$-$0.14.
\item Fig.~2. Isospin  0 $s$ -- wave \mbox{$\pi\pi$} phase shifts
calculated for the data
set~1 (solid line) and set~2 (dashed line). Arrows indicate the energy
range used
in the data fits. Data are from [28--31].
\item Fig.~3. Sum of the \mbox{$\pi\pi$} and \mbox{$K\overline{K}$}
phase shifts. Full circles denote data
set~1 \cite{cohen}, open circles -- data set~2 \cite{estabrooks}.
Lines and arrows as in Fig.~2.
\item Fig.~4. Inelasticity parameter $x=(1-\eta^{2})/4$.
Data are taken from \cite{cohen}. Lines and arrows as in Fig.~2.
\item Fig.~5. Structure of the $S_{\mbox{$\pi\pi$}}$ matrix element
in the complex
$z$ -- plane. Positions of poles (1 -- \mbox{$f_{0} (500)$}, 2 --
\mbox{$f_{0} (975)$}, 3 -- \mbox{$f_{0} (1400)$}) are
indicated by crosses and zeros by circles.
The roman numbers label the energy sheets. The bold
line shows the physical region and the rectangles on the imaginary
axis indicate the $S_{\mbox{$\pi\pi$}}$ cuts.
\end{itemize}

\end{document}